\documentclass[12pt]{article}
\usepackage{color}
\usepackage{amssymb,amsmath,comment}
\usepackage{graphicx}
\usepackage{braket}
\usepackage{epsfig}
\usepackage{here}
\usepackage{bm}
\graphicspath{{./Figures/}}
\newcommand{\ba}{\begin{align}}
\newcommand{\ea}{\end{align}}

\def\ov{\overline}
\def\nn{\nonumber}

\def\bea{\begin{eqnarray}}
\def\eea{\end{eqnarray}}
 \makeatletter
\def\alt{\mathrel{\mathpalette\gl@align<}}
\def\agt{\mathrel{\mathpalette\gl@align>}}
\def\gl@align#1#2{\lower.6ex\vbox{\baselineskip\z@skip\lineskip\z@
\ialign{$\m@th#1\hfil##\hfil$\crcr#2\crcr\sim\crcr}}} \makeatother
\renewcommand{\thefootnote}{\fnsymbol{footnote}}
\begin{document}
\begin{flushright}
\end{flushright}
\vspace*{1.0cm}

\begin{center}
\baselineskip 20pt 
{\Large\bf 
Moderately Suppressed Dimension-five Proton Decay 
\\
in a Flipped $SU(5)$ Model
}
\vspace{1cm}

{\large 
Naoyuki Haba$^a$ and Toshifumi Yamada$^b$
} \vspace{.5cm}

{\baselineskip 20pt \it
$^a$ Institute of Science and Engineering, Shimane University, Matsue 690-8504, Japan

$^b$ Department of Physics, Kyoto University, Kyoto, Kyoto 606-8502, Japan
}

\vspace{.5cm}

\vspace{1.5cm} {\bf Abstract} \end{center}

We study colored Higgsino-mediated proton decay (dimension-five proton decay)
 in a model based on the flipped $SU(5)$ GUT.
In the model, the GUT-breaking ${\bf10}$,~${\bf \ov{10}}$ fields
 have a GUT-scale mass term and gain VEVs through higher-dimensional operators,
 which induces an effective mass term
 between the color triplets in the ${\bf 5}$,~${\bf \bar{5}}$ Higgs fields
 that is not much smaller than the GUT scale.
This model structure gives rise to observable dimension-five proton decay, 
 and at the same time achieves moderate suppression on dimension-five proton decay that softens the tension with
 the current bound on $\Gamma(p\to K^+\bar{\nu})$.
We investigate the flavor dependence of the Wilson coefficients of the operators relevant to dimension-five proton decay,
 by relating them with diagonalized Yukawa couplings and CKM matrix components in MSSM,
 utilizing the fact that the GUT Yukawa couplings are in one-to-one correspondence with the MSSM Yukawa couplings
 in flipped models.
Then we numerically evaluate the Wilson coefficients,
 and predict the distributions of the ratios of the partial widths of various proton decay modes.

\thispagestyle{empty}

%\bigskip
\newpage
\renewcommand{\thefootnote}{\arabic{footnote}}
%\addtocounter{page}{-1}
\setcounter{footnote}{0}
%%%%%%%%%%%%%%%%%%%%%%%%%%
%\baselineskip 36pt
% Main body
%%%%%%%%%%%%%%%%%%%%%%%%%%
\baselineskip 18pt
%%%%%%%%%%%%%%%%%%%%%%%%%%

\section{Introduction}

Proton decay mediated by colored Higgsinos in supersymmetric grand unified theories (SUSY GUTs)~\cite{Weinberg:1981wj,Sakai:1981pk},
 called dimension-five proton decay,
 is a primary target in the proton decay searches at HyperKamiokande~\cite{Hyper-Kamiokande:2018ofw}, JUNO~\cite{JUNO:2015zny},
 and DUNE~\cite{DUNE:2015lol,DUNE:2020ypp}.
This is because the GUT gauge boson mass is predicted to be $\sim2\cdot10^{16}$~GeV in usual SUSY GUTs and 
 the corresponding partial width of GUT gauge boson-mediated proton decay is out of the sensitivity ranges of the above experiments
 (however, proton decay mediated by $({\bf 3},{\bf 2},1/6)$ gauge boson in SUSY $SO(10)$ GUT can be accessible~\cite{Haba:2019wwt}).
In non-SUSY GUTs, GUT gauge boson-mediated proton decay can be within the experimental reach, and one can even set upper bounds on the proton lifetime in some cases~\cite{Haba:2018vvu}.
Nevertheless, dependence on the choice of split $SU(5)$ multiplets that assist the gauge coupling unification, is inevitable.
In contrast, if Nature favors as light SUSY particles as possible
 (e.g., the SUSY particle mass spectrum is such that the $p\to K^+\bar{\nu}$ partial width is narrowly above the current experimental bound~\cite{Abe:2014mwa})
  for the naturalness of the electroweak scale,
 there is a great chance that dimension-five proton decay is observed in forthcoming experiments.
 (For study on proton decay and the LHC bounds on SUSY particle masses and the Higgs particle mass, 
 see Refs.~\cite{Ellis:2019fwf,Babu:2020ncc}.)
In this situation, it is important to study the flavor dependence of dimension-five proton decay in various SUSY GUT models, 
 and compare the partial widths of different decay modes, as in Ref.~\cite{Haba:2020bls},
 to bridge theory and proton decay search experiments.

In this paper, we focus on dimension-five proton decay in a model based on the SUSY flipped $SU(5)$ GUT~\cite{Derendinger:1983aj,Antoniadis:1987dx}
\footnote{
A non-SUSY model has first appeared in Ref.~\cite{Barr:1981qv}.
}.
Although the flipped $SU(5)$ GUT by itself cannot address the origin of the $U(1)$ hypercharge quantization,
 it has attractive features such as the realization of the doublet-triplet splitting, and the suppression of dimension-five proton decay
 that allows one to lower the sfermion mass without conflicting the current bound on the $p\to K^+\bar{\nu}$ decay.
In pervasive models of the SUSY flipped $SU(5)$ GUT, dimension-five proton decay is in effect totally suppressed, because the mass term between the color triplets
 in the ${\bf 5}$,~${\bf \bar{5}}$ Higgs fields
 and that between the color triplets in the GUT-breaking ${\bf10}$,~${\bf\ov{10}}$ fields,
 are at the soft SUSY breaking scale, not at the GUT scale.
However, if the GUT-breaking ${\bf10}$,~${\bf\ov{10}}$ fields
 are allowed to possess a GUT-scale mass term and gain vacuum expectation values (VEVs) through higher-dimensional operators,
 then the color triplets in the ${\bf 5}$,~${\bf \bar{5}}$ fields gain an effective mass term not much smaller than the GUT scale,
 which gives rise to observable dimension-five proton decays.
Meanwhile, the operators obtained after integrating out the colored Higgsinos
 can be $O(100)$ times suppressed compared to those in non-flipped models,
 which mitigates the tension with the current experimental bound on the $p\to K^+\bar{\nu}$ mode.
We materialize the above possibility in our model of the SUSY flipped $SU(5)$ GUT,
 and investigate the flavor dependence of dimension-five proton decay in the model.
Interestingly, since the GUT Yukawa couplings for the ${\bf 5}$,~${\bf \bar{5}}$ Higgs fields are in one-to-one correspondence with
 the Yukawa couplings of the minimal SUSY Standard Model (MSSM), we have a strong predictive power on the flavor structure of 
 the Wilson coefficients of the operators relevant to dimension-five proton decay.
We take advantage of the above feature
  and express the Wilson coefficients
  in terms of diagonalized Yukawa couplings and Cabibbo-Kobayashi-Maskawa (CKM) matrix components in MSSM, plus one unknown unitary matrix and several unknown phases.
Then we vary the unknown unitary matrix and phases and predict the distributions of the partial width ratios of different proton decay modes.

Previously, colored Higgsino-mediated proton decay in the SUSY flipped $SU(5)$ GUT has been studied in Ref.~\cite{Mehmood:2020irm}.
However, since the mass term between the color triplets of the ${\bf 5}$,~${\bf \bar{5}}$ Higgs fields 
 and that between the color triplets of the GUT-breaking ${\bf 10}$,~${\bf \bar{10}}$ fields
 are highly suppressed,
 only chirality non-flipping colored Higgsino exchange is considered, 
 unlike the present paper where we focus on chirality flipping colored Higgsino exchange.
For reference, GUT gauge boson-mediated proton decay in SUSY flipped $SU(5)$ GUT models has been investigated in
 Refs.\cite{Ellis:1993ks}-\cite{Hamaguchi:2020tet}.

This paper is organized as follows.
In Section~\ref{model}, we describe our model of the SUSY flipped $SU(5)$ GUT.
In Section~\ref{wilson}, we present the expressions for the Wilson coefficients of dimension-five operators obtained after integrating out colored Higgs fields and dimension-six operators obtained after integrating out the SUSY particles which contribute to proton decay.
In Section~\ref{wilson2}, we investigate the flavor dependence of the Wilson coefficients
 by relating them with diagonalized Yukawa couplings and CKM matrix in MSSM.
In Section~\ref{numerical}, we numerically evaluate the Wilson coefficients 
 using the values of diagonalized Yukawa couplings and CKM matrix based on experimental data,
 and randomly varying the remaining unknown parameters.
The results are presented as a prediction for the distributions of proton decay partial width ratios.
Section~\ref{summary} summarizes the paper.
\\

\section{Model}
\label{model}

We consider a model based on $SU(5)\times U(1)_X$ gauge group and the $Z_2$ matter parity.
The field content is as follows:
 three generations of chiral superfields in $({\bf 10},1)$, $({\bf \bar{5}},-3)$, $({\bf 1},5)$ representations of $SU(5)\times U(1)_X$ and with $Z_2$ matter parity $-1$, denoted by ${\bf 10}_1^i$, ${\bf \bar{5}}_{-3}^i$, ${\bf 1}_5^i$ where $i$ labels the three generations;
 chiral superfields in $({\bf 10},1)$,  $({\bf \overline{10}},-1)$ representations and with $Z_2$ matter parity $+1$, denoted by $H$, $\ov{H}$;
 chiral superfields in $({\bf 5},-2)$,  $({\bf \bar{5}},2)$ representations and with $Z_2$ matter parity $+1$, denoted by $h$, $\overline{h}$.
Additionally, we introduce a chiral superfield in $({\bf 24},0)$ representation with $Z_2$ matter parity $+1$, denoted by $\Sigma$,
 and three generations of chiral superfields in $({\bf 1},0)$ representation with $Z_2$ matter parity $-1$,
 denoted by $S^i$.
The field content is summarized in Table~\ref{content}.
\begin{table}[H]
\begin{center}
  \caption{Field content. $i$ labels the three generations.}
  \begin{tabular}{|c||c|c|c|} \hline
  chiral superfield & $SU(5)$ & $U(1)_X$ & $R$-parity \\ \hline
  ${\bf 10}_1^i$ & ${\bf 10}$                     & 1 & $-1$  \\
  ${\bf \bar{5}}_{-3}^i$ & ${\bf \bar{5}}$     & $-3$ & $-1$  \\
  ${\bf 1}_{5}^i$ & ${\bf 1}$                       & 5      & $-1$  \\ \hline
  $H$   & ${\bf 10}$                                  & 1 & $+1$  \\
  $\ov{H}$   & ${\bf \ov{10}}$                    & $-1$ & $+1$  \\
  $h$   & ${\bf 5}$                                  & $-2$ & $+1$  \\
  $\ov{h}$   & ${\bf \bar{5}}$                    & 2 & $+1$  \\ \hline
  $\Sigma$ & ${\bf 24}$                           & 0 & $+1$ \\ \hline
  $S^i$       &  ${\bf 1}$                            & 0 & $-1$  \\ \hline
  \end{tabular}
  \label{content}
  \end{center}
\end{table}
The fundamental superpotential reads
\begin{align}
W' &= (Y_d)_{ij}\, {\bf 10}_1^i\,  {\bf 10}_1^j\, h + (Y_u)_{ij}\, {\bf 10}_1^i\,  {\bf \bar{5}}_{-3}^j\, \overline{h}
+ (Y_e)_{ij}\, {\bf \bar{5}}_{-3}^i\, {\bf 1}_5^j\, h
\nn\\
&+ \mu_h\,h\overline{h} + \lambda\, HHh + \lambda'\,\ov{H}\,\ov{H}\,\ov{h} + M\,\ov{H}H 
\nn\\
&+ \kappa\,\Sigma\ov{H}H + \frac{1}{2}M_\Sigma \, \Sigma^2 + \frac{1}{3}\lambda_\Sigma \, \Sigma^3 
\nn\\
&+ (Y_S)_{ij} \, {\bf 10}_1^i \,S^j \,\ov{H} + \frac{1}{2}M_{S^i} \, S^i S^i.
\label{superpotential-pre}
\end{align}
Here we assume that there are no higher-dimensional operators at this stage.
The mass terms for the $\Sigma$ and $S^i$ fields are assumed to originate from some Planck-scale physics and 
 their masses $M_\Sigma$, $M_{S^i}$ are about the reduced Planck mass $M_P=2.4\cdot10^{18}$~GeV.
On the other hand, $\mu_h$ is at the soft SUSY breaking scale, while $M$ is about (GUT scale)$^2/M_P$, as shown later.
The origin of the hierarchy $|\mu_h|\ll |M|$ can be explained by introducing an $R$-symmetry 
 under which $H,\ov{H}$ and the matter superfields have $R$-charge 1 and $h,\ov{h}$ have 0
 and by assuming that the Planck-scale physics that gives rise to $M_\Sigma$, $M_{S^i}$ does not respect this $R$-symmetry.
By integrating out $\Sigma$ and $S^i$, we obtain the following effective superpotential at scales below $M_P$:
\begin{align}
W &= (Y_d)_{ij}\, {\bf 10}_1^i\,  {\bf 10}_1^j\, h + (Y_u)_{ij}\, {\bf 10}_1^i\,  {\bf \bar{5}}_{-3}^j\, \overline{h}
+ (Y_e)_{ij}\, {\bf \bar{5}}_{-3}^i\, {\bf 1}_5^j\, h
\nn\\
&+ \mu_h\,h\overline{h} + \lambda\, HHh + \lambda'\,\ov{H}\,\ov{H}\,\ov{h} + M\,\ov{H}H 
\nn\\
&- \frac{\kappa^2}{2M_\Sigma}\ov{H}H\ov{H}H - \frac{(Y_S)_{ik}(Y_S)_{jk}}{2M_{S^k}} (\ov{H}{\bf 10}_1^i)(\ov{H}{\bf 10}_1^j),
\label{superpotential}
\end{align}
 where $\ov{H}H\ov{H}H$ stands for the coupling where $SU(5)$ indices are summed cyclically
 and $(\ov{H}{\bf 10}_1^i)(\ov{H}{\bf 10}_1^j)$ the couplings where $SU(5)$ indices are summed separately in each bracket.
The $\ov{H}H\ov{H}H$ operator, obtained by integrating out $\Sigma$, triggers GUT breaking,
 while the $(\ov{H}{\bf 10}_1^i)(\ov{H}{\bf 10}_1^j)$ operators, obtained by integrating out $S^i$'s, generate the Majorana mass of the singlet neutrinos.
The isospin-doublet components of $h,\ov{h}$ gain mass only from the $\mu_h\,h\overline{h}$ term, while the color-triplet components additionally
 gain GUT-scale mass from the $\lambda\, HHh + \ov{\lambda}\,\ov{H}\,\ov{H}\,\ov{h}$ terms after GUT-breaking,
 which achieves the doublet-triplet splitting.
Note that the effective superpotential Eq.~(\ref{superpotential}) does not contain higher-dimensional operators
 giving rise to proton decay ${\bf 10}_1^i{\bf 10}_1^j{\bf 10}_1^k{\bf \bar{5}}_{-3}^\ell$;
proton decay occurs only through the exchange of colored Higgsinos, colored Higgs bosons and GUT gauge bosons
 (the contributions of the latter two can be neglected in the present model).

Let us write the Standard Model (SM) gauge-singlet, color-triplet and color-anti-triplet components of $H$ as 
 $N_H^c$, $Q_H$, $D^c_H$,
 and the SM gauge-singlet, color-anti-triplet and color-triplet components of $\ov{H}$ as 
 $\ov{N}_H^c$, $\ov{Q}_H$, $\ov{D}^c_H$, respectively.
$N_H^c,\ov{N}_H^c$ develop VEVs as
\bea
\langle N_H^c \rangle\langle \ov{N}_H^c \rangle = 2M\,M_\Sigma/\kappa^2, 
\ \ \ \ \ \ \ \ |\langle N_H^c \rangle| = |\langle \ov{N}_H^c \rangle|
\label{vev}
\eea
 and break $SU(5)\times U(1)_X$ into the SM gauge group.
The $Q_H,\ov{Q}_H$ fields are would-be Nambu-Goldstone modes eaten by the GUT gauge boson.
Along the GUT breaking, $h$ is decomposed into isospin-doublet $H_d$ and color-triplet $\ov{D}_h^c$,
 and $\ov{h}$ into isospin-doublet $H_u$ and color-anti-triplet $D_h^c$.
The $H_d$, $H_u$ fields are identified with the MSSM down-type and up-type Higgs fields, respectively.
The $\ov{D}_h^c$, $\ov{D}_H^c$, $D_h^c$, $D_H^c$ fields constitute the colored Higgs fields.
Along the GUT breaking, ${\bf 10}_1^i$, ${\bf \bar{5}}_{-3}^i$, ${\bf 1}_5^i$ fields are decomposed as
\bea
{\bf 10}_1^i \to Q^i, \, D^{c\,i}, \, N^{c\, i}, \ \ \ \ \ \ \ \ {\bf \bar{5}}_{-3}^i \to U^{c\,i},\, L^i, \ \ \ \ \ \ \ \ {\bf 1}_5^i \to E^{c\,i}
\nn
\eea
 where $Q^i$, $U^{c\,i}$, $D^{c\,i}$, $L^i$, $E^{c\,i}$, $N^{c\,i}$ are identified with the MSSM quark doublets, up-type quark singlets, down-type quark singlets,
 lepton doublets, charged leptons, and the singlet neutrinos, respectively.
Couplings $(Y_d)_{ij}$, $(Y_u)_{ij}$, $(Y_e)_{ij}$ are identified with the down-type quark, up-type quark and charged lepton Yukawa couplings of MSSM, respectively.

The tiny active neutrino mass is generated via the Type-1 seesaw mechanism~\cite{seesaw1}-\cite{seesaw4}.
The Majorana mass term for the singlet neutrinos is generated 
 from the second term of the third line of Eq.~(\ref{superpotential})
 along the GUT breaking as
\bea
W \ \supset \ -\frac{1}{2}N^{c\, i}N^{c\, j}\ \frac{(Y_S)_{ik}(Y_S)_{jk}}{2M_{S^k}}\langle \ov{N}_H^c \rangle^2.
\eea
Since the Dirac mass for $N^{c\, i}$ and $L^j$ is given by $(Y_u)_{ij} v_u/\sqrt{2}$
 with $v_u=\sin\beta\cdot246$~GeV,
 the active neutrino mass matrix is obtained as
\bea
M_\nu \ = \ Y_u^T Y_S^{T\,-1}M_SY_S^{-1} Y_u \ \frac{v_u^2}{\langle \ov{N}_H^c \rangle^2}
\eea
 where $M_S={\rm diag}(M_{S^1},\,M_{S^2},\,M_{S^3})$.
We estimate the scale of the components of $Y_S$.
Consider a special case where $M_S=M_P\,{\rm diag}(1,1,1)$ and $Y_u$ is given by
\bea
Y_u = {\rm diag}(y_u,\,y_c,\,y_t)\,U_{MNS}^\dagger
\eea
 in the basis where the lepton-doublet components of ${\bf \bar{5}}_{-3}^i$ diagonalize the charged lepton mass matrix.
Here $U_{MNS}$ denotes Maki-Nakagawa-Sakata matrix~\cite{mns}.
In this case, we get
\bea
Y_S = {\rm diag}(y_u/\sqrt{m_1},\,y_c/\sqrt{m_2},\,y_t/\sqrt{m_3})\ \frac{v_u \sqrt{M_P}}{\langle \ov{N}_H^c \rangle},
\eea
 where $m_1,m_2,m_3$ are the active neutrino masses.
For $|\langle \ov{N}_H^c \rangle|\simeq 3\cdot10^{16}$~GeV, $v_u\simeq246$~GeV,
 and for the normal hierarchy with $m_2\gg m_1$,
 the following numerical values reproduce the measured neutrino mass differences:
\bea
Y_S = {\rm diag}(a,\,0.006,\,0.9) \ \ \ \ \ {\rm with} \ a>10^{-4}.
\eea
We expect that in general cases suppression of some components of $Y_S$ at $O(10^{-3})$
 suffices to reproduce the measured neutrino mass differences.
\\

\section{Wilson Coefficients Contributing to Proton Decay}
\label{wilson}

We focus on proton decay mediated by colored Higgsinos.
The mass matrix for the colored Higgs fields reads
\bea
W \supset    \begin{pmatrix} % or pmatrix or bmatrix or Bmatrix or ...
      D_h^c & D_H^c \\
   \end{pmatrix}
      {\cal M}_{H_C}
         \begin{pmatrix} % or pmatrix or bmatrix or Bmatrix or ...
            \ov{D}_h^c \\
            \ov{D}_H^c \\
         \end{pmatrix}, \ \ \ \ \ {\cal M}_{H_C}=\begin{pmatrix} % or pmatrix or bmatrix or Bmatrix or ...
         \mu_h & \lambda' \langle \ov{N}_H^c\rangle \\
         \lambda \langle N_H^c\rangle & M \\
      \end{pmatrix}.
      \label{coloredhiggsmass}
\eea
The colored Higgs fields couple to the matter fields as
\bea
W \supset\ (Y_d)_{ij}Q^iQ^j\,\ov{D}_h^c + (Y_d)_{ij}D^{c\,i}N^{c\, j}\,\ov{D}_h^c + (Y_u)_{ij}Q^iL^j\,D_h^c + (Y_u)_{ij}D^{c\,i}U^{c\,j}\,D_h^c + (Y_e)_{ij}U^{c\,i}E^{c\,j}\,\ov{D}_h^c.
\eea

Integrating out the colored Higgs fields, we obtain the following dimension-five operators responsible for proton decay:
\bea
-W_5 \ = \ \frac{1}{2}C_{5L}^{ijkl}\,(Q^k Q^l)(Q^i L^j) + C_{5R}^{ijkl}\,E^{c\,k} U^{c\,l} U^{c\,i} D^{c\,j}
\label{dim5}
\eea
 where in the first term isospin indices are summed in each bracket, and the Wilson coefficients satisfy
\bea
C_{5L}^{ijkl}(\mu=\mu_{H_C}) &=&  ({\cal M}_{H_C}^{-1})_{11}\left.\left\{(Y_d)_{kl}(Y_u)_{ij} - \frac{1}{2}(Y_d)_{li}(Y_u)_{kj} - \frac{1}{2}(Y_d)_{ik}(Y_u)_{lj}\right\}\right|_{\mu=\mu_{H_C}},
\nn\\
C_{5R}^{ijkl}(\mu=\mu_{H_C}) &=&  ({\cal M}_{H_C}^{-1})_{11}\left.\left\{(Y_e)_{lk}(Y_u)_{ji} - (Y_e)_{ik}(Y_u)_{jl}\right\}\right|_{\mu=\mu_{H_C}}
\eea
 where $\mu$ denotes the renormalization scale and $\mu_{H_C}$ is about the colored Higgs mass eigenvalues.
The effective inverse of the $D_h^c$, $\ov{D}_h^c$ fields, $({\cal M}_{H_C}^{-1})_{11}$, is obtained from Eq.~(\ref{coloredhiggsmass}) as
\bea
({\cal M}_{H_C}^{-1})_{11} = -\frac{M}{\lambda\lambda'\langle N_H^c\rangle\langle \ov{N}_H^c\rangle} = 
-\frac{\kappa^2}{2\lambda\lambda' M_\Sigma},
\label{mhcinverse}
\eea
 where we have used the fact that $\mu_h$ is negligible compared to the GUT-scale, and used Eq.~(\ref{vev}) in the second equality.
When $|\kappa^2/(2\lambda\lambda')|=1$ and $M_\Sigma=M_P$,
 the Wilson coefficients $C_{5L}^{ijkl},C_{5R}^{ijkl}$ are about 100 times suppressed compared to non-flipped models, where
 $({\cal M}_{H_C}^{-1})_{11}$ is given by the inverse of the colored Higgs mass $\sim 2\cdot 10^{16}$~GeV.
The resulting $10^4$ suppression on proton decay partial widths allows the model to evade 
 the current stringent experimental bound on the $p\to K^+\bar{\nu}$ decay without enormously raising sfermion masses.
Still, the suppression is not strong, and leaves the possibility of observing proton decay in near-future experiments.

Integrating out the SUSY particles, we obtain the following operators responsible for proton decay:
\bea
-{\cal L}_6=  C_{LL}^{ijkl}(\psi_{u_L}^k\psi_{d_L}^l)(\psi_{d_L}^i\psi_{\nu_L}^j) + \ov{C}_{LL}^{ijkl}(\psi_{u_L}^k\psi_{d_L}^l)(\psi_{u_L}^i\psi_{e_L}^j)
+ C_{RL}^{ijkl}(\psi_{\nu_L}^k\psi_{d_L}^l)(\psi_{u_R^c}^i\psi_{d_R^c}^j)
\label{dim6}
\eea
 where $\psi$ denotes a SM Weyl spinor and spinor index is summed in each bracket. 
Those Wilson coefficients which contribute to proton decay are 
 $C_{LL}^{d\alpha\,ud}, C_{LL}^{s\alpha\,ud}$, $C_{LL}^{d\alpha\,us}$, $\ov{C}_{LL}^{u\beta\,ud}$, $\ov{C}_{LL}^{u\beta\,us}$,
 $C_{RL}^{ud\,\tau d}$, $C_{RL}^{ud\,\tau s}$, $C_{RL}^{us\,\tau d}$
 with $\alpha=e,\mu,\tau$ and $\beta=e,\mu$.
They satisfy, at the soft SUSY breaking scale $\mu=\mu_{\rm SUSY}$,
\footnote{
  When writing $C_{5L}^{s\alpha\, ud}$, we mean that $Q_i$ is in the flavor basis where the down-type quark Yukawa coupling is
  diagonal and that the down-type quark component of $Q_i$ is exactly $s$ quark (the up-type quark component of $Q_i$ is a mixture of $u,c,t$).
  Likewise, $Q_k$ is in the flavor basis where the up-type quark Yukawa coupling is
  diagonal and its up-type component is exactly $u$ quark, 
  and $Q_l$ is in the flavor basis where the down-type quark Yukawa coupling is
  diagonal and its down-type quark component is exactly $d$ quark.
  The same rule applies to $C_{5L}^{u\alpha \,ds}$ and others.
  }
\begin{align}
C_{LL}^{d\alpha\,ud}(\mu_{\rm SUSY})= \frac{M_{\widetilde{W}}}{m_{\tilde{q}}^2}
\, {\cal F}\left(|M_{\widetilde{W}}|^2/m_{\tilde{q}}^2,~m_{\tilde{\ell}^\alpha}^2/m_{\tilde{q}}^2
\right)\,
g_2^2\left(C_{5L}^{d\alpha\,ud}-C_{5L}^{u\alpha\, dd}\right)\vert_{\mu=\mu_{\rm SUSY}},
\label{ll1}
\\
C_{LL}^{s\alpha\,ud}(\mu_{\rm SUSY})= \frac{M_{\widetilde{W}}}{m_{\tilde{q}}^2}
\, {\cal F}\left(|M_{\widetilde{W}}|^2/m_{\tilde{q}}^2,~m_{\tilde{\ell}^\alpha}^2/m_{\tilde{q}}^2
\right)\,
g_2^2\left(C_{5L}^{s\alpha\,ud}-C_{5L}^{u\alpha\, ds}\right)\vert_{\mu=\mu_{\rm SUSY}},
\\
C_{LL}^{d\alpha\,us}(\mu_{\rm SUSY})= \frac{M_{\widetilde{W}}}{m_{\tilde{q}}^2}
\,{\cal F}\left(|M_{\widetilde{W}}|^2/m_{\tilde{q}}^2,~m_{\tilde{\ell}^\alpha}^2/m_{\tilde{q}}^2
\right)\,
g_2^2\left(C_{5L}^{d\alpha\, us}-C_{5L}^{u\alpha\, ds}\right)\vert_{\mu=\mu_{\rm SUSY}},
\\
\ov{C}_{LL}^{u\beta\,ud}(\mu_{\rm SUSY})= \frac{M_{\widetilde{W}}}{m_{\tilde{q}}^2}
\,{\cal F}\left(|M_{\widetilde{W}}|^2/m_{\tilde{q}}^2,~m_{\tilde{\ell}^\beta}^2/m_{\tilde{q}}^2
\right)\,
g_2^2\left(-C_{5L}^{u\beta\, ud}+C_{5L}^{d\beta\, uu}\right)\vert_{\mu=\mu_{\rm SUSY}},
\\
\ov{C}_{LL}^{u\beta\,us}(\mu_{\rm SUSY})= \frac{M_{\widetilde{W}}}{m_{\tilde{q}}^2}
\,{\cal F}\left(|M_{\widetilde{W}}|^2/m_{\tilde{q}}^2,~m_{\tilde{\ell}^\beta}^2/m_{\tilde{q}}^2
\right)\,
g_2^2\left(-C_{5L}^{u\beta\, us}+C_{5L}^{s\beta\, uu}\right)\vert_{\mu=\mu_{\rm SUSY}},
\\
C_{RL}^{ud\,\tau d}(\mu_{\rm SUSY})= \frac{\mu_h}{m_{\tilde{t}_R}^2}\,
{\cal \tilde{F}}\left(|\mu_h|^2/m_{\tilde{t}_R}^2,~m_{\tilde{\tau}_R}^2/m_{\tilde{t}_R}^2
\right)
\ (V^{\rm ckm}_{td})^*\, y_t y_\tau
\,C_{5R}^{ud\tau t}\vert_{\mu=\mu_{\rm SUSY}},
\\
C_{RL}^{ud\,\tau s}(\mu_{\rm SUSY})= \frac{\mu_h}{m_{\tilde{t}_R}^2}\,
{\cal \tilde{F}}\left(|\mu_h|^2/m_{\tilde{t}_R}^2,~m_{\tilde{\tau}_R}^2/m_{\tilde{t}_R}^2
\right)
\ (V^{\rm ckm}_{ts})^*\, y_t y_\tau
\,C_{5R}^{ud\tau t}\vert_{\mu=\mu_{\rm SUSY}},
\\
C_{RL}^{us\,\tau d}(\mu_{\rm SUSY})= \frac{\mu_h}{m_{\tilde{t}_R}^2}\,
{\cal \tilde{F}}\left(|\mu_h|^2/m_{\tilde{t}_R}^2,~m_{\tilde{\tau}_R}^2/m_{\tilde{t}_R}^2
\right)
\ (V^{\rm ckm}_{td})^*\, y_t y_\tau
\,C_{5R}^{us\tau t}\vert_{\mu=\mu_{\rm SUSY}}.
\label{rl2}
\end{align}
Here ${\cal F},{\cal \tilde{F}}$ are loop functions given by
 ${\cal F}(x,y) =\frac{1}{x-y}(\frac{x}{1-x}\log x - \frac{y}{1-y}\log y)/16\pi^2 + \frac{1}{x-1}(\frac{x}{1-x}\log x+1)/16\pi^2$
  and ${\cal \tilde{F}}(x,y) =\frac{1}{x-y}(\frac{x}{1-x}\log x - \frac{y}{1-y}\log y)/16\pi^2$,
  and $M_{\widetilde{W}}$, $\mu_h$, $m_{\tilde{q}}$, $m_{\tilde{\ell}^\alpha}$, $m_{\tilde{t}_R}$, $m_{\tilde{\tau}_R}$ respectively denote
 the Wino mass, $\mu$-term, mass of 1st and 2nd generation isospin-doublet squarks (assumed degenerate), 
 mass of isospin-doublet slepton of flavor $\alpha$,
 mass of isospin-singlet top squark, and mass of isospin-singlet tau slepton
 (mixings between isospin-doublet and singlet sfermions are neglected).
$y_t$, $y_\tau$, $V^{\rm ckm}$ denote the top quark and tau lepton Yukawa couplings and CKM matrix,
 respectively.

Finally, $C_{LL}^{d\alpha\,ud}, C_{LL}^{s\alpha\,ud}$, $C_{LL}^{d\alpha\,us}$, $\ov{C}_{LL}^{u\beta\,ud}$, $\ov{C}_{LL}^{u\beta\,us}$,
 $C_{RL}^{ud\,\tau d}$, $C_{RL}^{ud\,\tau s}$, $C_{RL}^{us\,\tau d}$ at a hadronic scale $\mu=\mu_{\rm had}$ determine
 proton decay amplitudes.
\\

\section{Flavor Dependence of Wilson Coefficients}
\label{wilson2}

We investigate the flavor dependence of the Wilson coefficients 
 $C_{LL}^{d\alpha\,ud}, C_{LL}^{s\alpha\,ud}$, $C_{LL}^{d\alpha\,us}$, $\ov{C}_{LL}^{u\beta\,ud}$, $\ov{C}_{LL}^{u\beta\,us}$,
 $C_{RL}^{ud\,\tau d}$, $C_{RL}^{ud\,\tau s}$, $C_{RL}^{us\,\tau d}$ ($\alpha=e,\mu,\tau$ and $\beta=e,\mu$)
 by relating them with diagonalized Yukawa couplings and the CKM matrix in MSSM.
We can write
\begin{align}
C_{LL}^{d\alpha\,ud}(\mu_{\rm had}) \ &= \ C_{LL\alpha}^0 \left.\left\{(Y_d)_{u_L d_L}(Y_u)_{d_L\alpha_L} - (Y_d)_{d_L d_L}(Y_u)_{u_L \alpha_L}\right\}\right|_{\mu=\mu_{H_C}},
\label{cll1}\\
C_{LL}^{s\alpha\,ud}(\mu_{\rm had}) \ &= \ C_{LL\alpha}^0 \left.\left\{(Y_d)_{u_L d_L}(Y_u)_{s_L\alpha_L} - (Y_d)_{d_L s_L}(Y_u)_{u_L \alpha_L}\right\}\right|_{\mu=\mu_{H_C}},
\label{cll2}\\
C_{LL}^{d\alpha\,us}(\mu_{\rm had}) \ &= \ C_{LL\alpha}^0 \left.\left\{(Y_d)_{u_L s_L}(Y_u)_{d_L\alpha_L} - (Y_d)_{d_L s_L}(Y_u)_{u_L \alpha_L}\right\}\right|_{\mu=\mu_{H_C}},
\label{cll3}\\
\ov{C}_{LL}^{u\beta\,ud}(\mu_{\rm had}) \ &= \ C_{LL\beta}^0 \left.\left\{(Y_d)_{u_L u_L}(Y_u)_{d_L\beta_L} - (Y_d)_{u_L d_L}(Y_u)_{u_L \beta_L}\right\}\right|_{\mu=\mu_{H_C}},
\label{cll4}\\
\ov{C}_{LL}^{u\beta\,us}(\mu_{\rm had}) \ &= \ C_{LL\beta}^0 \left.\left\{(Y_d)_{u_L u_L}(Y_u)_{s_L\beta_L} - (Y_d)_{u_L s_L}(Y_u)_{u_L \beta_L}\right\}\right|_{\mu=\mu_{H_C}},
\label{cll5}\\
C_{RL}^{ud\,\tau d}(\mu_{\rm had}) \ &= \ C_{RL}^0 \ (V^{\rm ckm}_{td})^*|_{\mu=\mu_{\rm SUSY}} \left.\left\{(Y_e)_{t_R \tau_R}(Y_u)_{d_Ru_R} - (Y_e)_{u_R \tau_R}(Y_u)_{d_Rt_R}\right\}\right|_{\mu=\mu_{H_C}},
\label{crl}\\
C_{RL}^{ud\,\tau s}(\mu_{\rm had}) \ &= \ C_{RL}^0 \ (V^{\rm ckm}_{ts})^*|_{\mu=\mu_{\rm SUSY}} \left.\left\{(Y_e)_{t_R \tau_R}(Y_u)_{d_Ru_R} - (Y_e)_{u_R \tau_R}(Y_u)_{d_Rt_R}\right\}\right|_{\mu=\mu_{H_C}},
\label{crl1}\\
C_{RL}^{us\,\tau d}(\mu_{\rm had}) \ &= \ C_{RL}^0 \ (V^{\rm ckm}_{td})^*|_{\mu=\mu_{\rm SUSY}} \left.\left\{(Y_e)_{t_R \tau_R}(Y_u)_{s_Ru_R} - (Y_e)_{u_R \tau_R}(Y_u)_{s_Rt_R}\right\}\right|_{\mu=\mu_{H_C}}.
\label{crl2}
\end{align}
Here $C_{LL\alpha}^0$ ($\alpha=e,\mu,\tau$), $C_{RL}^0$ are defined as
\begin{align}
C_{LL\alpha}^0 &= \frac{3}{2}({\cal M}_{H_C}^{-1})_{11}\ A_{L\alpha}^{\rm MSSM} \ \frac{M_{\widetilde{W}}}{m_{\tilde{q}}^2}{\cal F}\left(|M_{\widetilde{W}}|^2/m_{\tilde{q}}^2,~m_{\tilde{\ell}^\alpha}^2/m_{\tilde{q}}^2
\right)\, A_{LL}^{\rm SM} \ \ g_2^2|_{\mu=\mu_{\rm SUSY}},
\label{cll0}\\
C_{RL}^0 &= ({\cal M}_{H_C}^{-1})_{11}\ A_{Rt\tau}^{\rm MSSM} \ \frac{\mu_h}{m_{\tilde{t}_R}^2}{\cal \tilde{F}}\left(|\mu_h|^2/m_{\tilde{t}_R}^2,~m_{\tilde{\tau}_R}^2/m_{\tilde{t}_R}^2
\right)\, A_{RL}^{\rm SM} \ \ y_t y_\tau|_{\mu=\mu_{\rm SUSY}}
\label{crl0}
\end{align}
 where $A_{L\alpha}^{\rm MSSM}$ represents renormalization group (RG) corrections for $C_{5L}^{ijkl}$ in MSSM 
 that involve three light-flavor (1st and 2nd generations) quarks and one lepton of flavor $\alpha$,
 and $A_{Rt\tau}^{\rm MSSM}$ those for $C_{5R}^{ijkl}$ in MSSM 
 that involve two light-flavor quarks, one top quark and one tau lepton.
$A_{LL}^{\rm SM}$ and $A_{RL}^{\rm SM}$ respectively represent RG corrections for $C_{LL}^{ijkl},\ov{C}_{LL}^{ijkl}$ and $C_{RL}^{ijkl}$ 
 in SM involving no top quark.
When defining $A_{L\alpha}^{\rm MSSM}$, $A_{Rt\tau}^{\rm MSSM}$, $A_{LL}^{\rm SM}$, $A_{RL}^{\rm SM}$,
 we neglect the 1st and 2nd generation Yukawa couplings in MSSM, and the Yukawa couplings other than the top quark's in SM.
 (Then we have $A_{Le}^{\rm MSSM}=A_{L\mu}^{\rm MSSM}$, but we adhere to the redundant notation with $A_{L\alpha}^{\rm MSSM}$
 for brevity.)

The Yukawa coupling components in Eqs.~(\ref{cll1})-(\ref{crl2}) are related to diagonalized Yukawa couplings and CKM matrix components as follows:
Since $Y_d$ is a symmetric matrix, and since $Y_d$ by definition satisfies $(Y_d)_{d_Ld_R}=y_d$, $(Y_d)_{s_Ls_R}=y_s$, $(Y_d)_{b_Lb_R}=y_b$,
 $(Y_d)_{d_Ls_R}=(Y_d)_{d_Lb_R}=(Y_d)_{s_Ld_R}=(Y_d)_{s_Lb_R}=(Y_d)_{b_Ld_R}=(Y_d)_{b_Ls_R}=0$, we have
\bea
&&(Y_d)_{d_L d_L} = y_d \,e^{i\,\phi_1}, \ \ \ \ \ (Y_d)_{s_L s_L} = y_s \,e^{i\,\phi_2}, \ \ \ \ \ (Y_d)_{b_L b_L} = y_b \,e^{i\,\phi_3},
\nn\\
&&(Y_d)_{d_L s_L} = (Y_d)_{s_L b_L} = (Y_d)_{b_L d_L}=(Y_d)_{s_L d_L} = (Y_d)_{b_L s_L} = (Y_d)_{d_L b_L}=0
\label{ydcomponents}
\eea
 where $y_d,y_s,y_b$ denote the diagonalized Yukawa couplings for the down, strange and bottom quarks, respectively,
 and $\phi_1,\phi_2,\phi_3$ are unknown phases.
Combining Eq.~(\ref{ydcomponents}) with the definition of the CKM matrix, we get
\begin{align}
(Y_d)_{u_Ld_L} &= (V^{\rm ckm}_{ud})^*\,(Y_d)_{d_L d_L}=(V^{\rm ckm}_{ud})^*\,y_d \,e^{i\,\phi_1},
\nn\\
(Y_d)_{u_Ls_L} &= (V^{\rm ckm}_{us})^*\,(Y_d)_{s_L s_L}=(V^{\rm ckm}_{us})^*\,y_s \,e^{i\,\phi_2},
\nn\\
(Y_d)_{u_Lu_L} &= 
(V^{\rm ckm}_{ud})^*\,(Y_d)_{d_L d_L}(V^{\rm ckm}_{ud})^* + (V^{\rm ckm}_{us})^*\,(Y_d)_{s_L s_L}(V^{\rm ckm}_{us})^* + (V^{\rm ckm}_{ub})^*\,(Y_d)_{b_L b_L}(V^{\rm ckm}_{ub})^*
\nn\\
&=(V^{\rm ckm}_{ud})^{*2}\,y_d \,e^{i\,\phi_1}+(V^{\rm ckm}_{us})^{*2}\,y_s \,e^{i\,\phi_2}+(V^{\rm ckm}_{ub})^{*2}\,y_b \,e^{i\,\phi_3}
\label{ydcomponents2}
\end{align}
 where $V^{\rm ckm}_{ij}$ denote components of the CKM matrix.
From the definition of $Y_u$ and the CKM matrix, we get, for $\alpha=e,\mu,\tau$,
\begin{align}
(Y_u)_{u_L\alpha_L} &= (Y_u)_{u_Lu_R} U_{u_R \alpha_L} = y_u\,U_{u_R \alpha_L},
\nn\\
(Y_u)_{d_L\alpha_L} &= V^{\rm ckm}_{ud} (Y_u)_{u_Lu_R} U_{u_R \alpha_L} + V^{\rm ckm}_{cd} (Y_u)_{c_Lc_R} U_{c_R \alpha_L}
+ V^{\rm ckm}_{td} (Y_u)_{t_Lt_R} U_{t_R \alpha_L}
\nn\\
&=V^{\rm ckm}_{ud}\,y_u \,U_{u_R \alpha_L} + V^{\rm ckm}_{cd} \, y_c \,U_{c_R \alpha_L}
+ V^{\rm ckm}_{td} \,y_t \,U_{t_R \alpha_L},
\nn\\
(Y_u)_{s_L\alpha_L}
&=V^{\rm ckm}_{us}\,y_u \,U_{u_R \alpha_L} + V^{\rm ckm}_{cs} \, y_c \,U_{c_R \alpha_L}
+ V^{\rm ckm}_{ts} \,y_t \,U_{t_R \alpha_L}
\label{yucomponents}
\end{align}
 where $y_u,y_c,y_t$ denote the diagonalized Yukawa couplings for the up, charm and top quarks, respectively,
 and $U_{ij}$ is a component of an unknown unitary matrix $U$ that transforms the flavor basis of ${\bf \bar{5}}_{-3}^i$'s as
\bea
   \begin{pmatrix} % or pmatrix or bmatrix or Bmatrix or ...
       {\bf \bar{5}}_{-3}^{u_R} \\
       {\bf \bar{5}}_{-3}^{c_R} \\
       {\bf \bar{5}}_{-3}^{t_R} \\
   \end{pmatrix}
   = U
   \begin{pmatrix} % or pmatrix or bmatrix or Bmatrix or ...
       {\bf \bar{5}}_{-3}^{e_L} \\
       {\bf \bar{5}}_{-3}^{\mu_L} \\
       {\bf \bar{5}}_{-3}^{\tau_L} \\
   \end{pmatrix}.
 \label{udefinition}
\eea
Combining the definition of $Y_e$ with Eq.~(\ref{udefinition}), we get
\begin{align}
(Y_e)_{t_R\tau_R} &= U^*_{t_R\tau_L}y_\tau, \ \ \ \ \ \ (Y_e)_{u_R\tau_R} = U^*_{u_R\tau_L}y_\tau.
\label{yecomponents}
\end{align}
Eq.~(\ref{ydcomponents}) gives that the flavor basis of ${\bf 10}_1^i$ is transformed as
\bea
   \begin{pmatrix} % or pmatrix or bmatrix or Bmatrix or ...
       {\bf 10}_1^{d_R} \\
       {\bf 10}_1^{s_R} \\
       {\bf 10}_1^{b_R} \\
   \end{pmatrix}
   =    \begin{pmatrix} % or pmatrix or bmatrix or Bmatrix or ...
         e^{i\,\phi_1} & 0 & 0\\
         0 & e^{i\,\phi_2} & 0\\
         0 & 0 & e^{i\,\phi_3}\\
      \end{pmatrix}
   \begin{pmatrix} % or pmatrix or bmatrix or Bmatrix or ...
       {\bf 10}_1^{d_L} \\
       {\bf 10}_1^{s_L} \\
       {\bf 10}_1^{b_L} \\
   \end{pmatrix}.
\label{basisof10}
\eea
Combining Eq.~(\ref{basisof10}) with the definition of CKM matrix, we get
\begin{align}
(Y_u)_{d_Ru_R} &= e^{-i\,\phi_1}\,(Y_u)_{d_Lu_R} = e^{-i\,\phi_1}\,V^{\rm ckm}_{ud}\,y_u,
\ \ \ \ \ \ 
(Y_u)_{d_Rt_R} = e^{-i\,\phi_1}\,(Y_u)_{d_Lt_R} = e^{-i\,\phi_1}\,V^{\rm ckm}_{td}\,y_t,
\nn\\
(Y_u)_{s_Ru_R} &= e^{-i\,\phi_2}\,(Y_u)_{s_Lu_R} = e^{-i\,\phi_2}\,V^{\rm ckm}_{us}\,y_u,
\ \ \ \ \ \ 
(Y_u)_{s_Rt_R} = e^{-i\,\phi_2}\,(Y_u)_{s_Lt_R} = e^{-i\,\phi_2}\,V^{\rm ckm}_{ts}\,y_t.
\label{yucomponents2}
\end{align}
Inserting Eqs.~(\ref{ydcomponents})-(\ref{yucomponents}),(\ref{yecomponents}),(\ref{yucomponents2}) into Eqs.~(\ref{cll1})-(\ref{crl2}),
 we obtain
\begin{align}
C_{LL}^{d\alpha\,ud}(\mu_{\rm had})/C_{LL\alpha}^0
 &=  e^{i\,\phi_1}\left\{ (V^{\rm ckm}_{ud})^*y_d
(V^{\rm ckm}_{ud}\,y_u \,U_{u_R \alpha_L} + V^{\rm ckm}_{cd} \, y_c \,U_{c_R \alpha_L} + V^{\rm ckm}_{td} \,y_t \,U_{t_R \alpha_L}) -y_d\,y_uU_{u_R\alpha_L}\right\}
\label{cll1-2}
\\
C_{LL}^{s\alpha\,ud}(\mu_{\rm had})/C_{LL\alpha}^0 &= e^{i\,\phi_1}\, (V^{\rm ckm}_{ud})^*\,y_d\,
(V^{\rm ckm}_{us}\,y_u \,U_{u_R \alpha_L} + V^{\rm ckm}_{cs} \, y_c \,U_{c_R \alpha_L} + V^{\rm ckm}_{ts} \,y_t \,U_{t_R \alpha_L})
\label{cll2-2}
\\
C_{LL}^{d\alpha\,us}(\mu_{\rm had})/C_{LL\alpha}^0 &= e^{i\,\phi_2}\, (V^{\rm ckm}_{us})^*\,y_s\,
(V^{\rm ckm}_{ud}\,y_u \,U_{u_R \alpha_L} + V^{\rm ckm}_{cd} \, y_c \,U_{c_R \alpha_L} + V^{\rm ckm}_{td} \,y_t \,U_{t_R \alpha_L})
\label{cll3-2}
\\
\ov{C}_{LL}^{u\beta\,ud}(\mu_{\rm had})/C_{LL\beta}^0 &=  \left\{
\left((V^{\rm ckm}_{ud})^{*2}\,y_d \,e^{i\,\phi_1}+(V^{\rm ckm}_{us})^{*2}\,y_s \,e^{i\,\phi_2}+(V^{\rm ckm}_{ub})^{*2}\,y_b \,e^{i\,\phi_3}\right)\right.
\nn\\
&\ \ \ \ \ \ \ \ \ \ \ \ \ \ \ \ \ \ \ \ \ \ \ \ \ \ \ \ \ \ \ \times\left(V^{\rm ckm}_{ud}\,y_u \,U_{u_R \beta_L} + V^{\rm ckm}_{cd} \, y_c \,U_{c_R \beta_L} + V^{\rm ckm}_{td} \,y_t \,U_{t_R \beta_L}\right)
\nn\\
&\left.\ \ \ \ \ \ \ \ \ \ \ \ -(V^{\rm ckm}_{ud})^*\,y_d \,e^{i\,\phi_1}\,y_u\,U_{u_R \beta_L}\right\},
\label{cll4-2}
\\
\ov{C}_{LL}^{u\beta\,us}(\mu_{\rm had})/C_{LL\beta}^0 &=  \left\{
\left((V^{\rm ckm}_{ud})^{*2}\,y_d \,e^{i\,\phi_1}+(V^{\rm ckm}_{us})^{*2}\,y_s \,e^{i\,\phi_2}+(V^{\rm ckm}_{ub})^{*2}\,y_b \,e^{i\,\phi_3}\right)\right.
\nn\\
&\ \ \ \ \ \ \ \ \ \ \ \ \ \ \ \ \ \ \ \ \ \ \ \ \ \ \ \ \ \ \ \times\left(V^{\rm ckm}_{us}\,y_u \,U_{u_R \beta_L} + V^{\rm ckm}_{cs} \, y_c \,U_{c_R \beta_L} + V^{\rm ckm}_{ts} \,y_t \,U_{t_R \beta_L}\right)
\nn\\
&\left.\ \ \ \ \ \ \ \ \ \ \ \ -(V^{\rm ckm}_{us})^*\,y_s \,e^{i\,\phi_2}\,y_u\,U_{u_R \beta_L}\right\},
\label{cll5-2}
\\
C_{RL}^{ud\,\tau d}(\mu_{\rm had})/C_{RL}^0 &=  e^{-i\,\phi_1}(\ov{V}_{td}^{\rm ckm})^*\,y_\tau\,
(U^*_{t_R\tau_L}V_{ud}^{\rm ckm}y_u - U^*_{u_R\tau_L}V_{td}^{\rm ckm}y_t),
\label{crl1-0}
\\
C_{RL}^{ud\,\tau s}(\mu_{\rm had})/C_{RL}^0 &=  e^{-i\,\phi_1}(\ov{V}_{ts}^{\rm ckm})^*\,y_\tau\,
(U^*_{t_R\tau_L}V_{ud}^{\rm ckm}y_u - U^*_{u_R\tau_L}V_{td}^{\rm ckm}y_t),
\label{crl1-2}
\\
C_{RL}^{us\,\tau d}(\mu_{\rm had})/C_{RL}^0 &= e^{-i\,\phi_2}(\ov{V}_{td}^{\rm ckm})^*\,y_\tau\,
(U^*_{t_R\tau_L}V_{us}^{\rm ckm}y_u - U^*_{u_R\tau_L}V_{ts}^{\rm ckm}y_t),
\label{crl2-2}
\end{align}
 where the diagonalized Yukawa couplings and CKM matrix components are evaluated at scale $\mu=\mu_{H_C}$,
 except for $\ov{V}_{ts}^{\rm ckm}$, $\ov{V}_{td}^{\rm ckm}$, which are evaluated at $\mu=\mu_{\rm SUSY}$.
The right-hand sides of Eqs.~(\ref{cll1-2})-(\ref{crl2-2}) contain diagonalized Yukawa couplings and components of the CKM matrix,
 and mostly determine the order of magnitude of each Wilson coefficient.
\\

\section{Numerical Analysis}
\label{numerical}

We numerically evaluate the right-hand sides of Eqs.~(\ref{cll1-2})-(\ref{crl2-2})
 by randomly varying the unknown unitary matrix $U$ and unknown phases $\phi_1,\phi_2,\phi_3$.
The result is presented in the form of the ratios of the proton decay partial widths below,
\bea
&&\Gamma(p\to \pi^+ \bar{\nu}) = \sum_{\alpha=e,\mu,\tau}\Gamma(p\to \pi^+ \bar{\nu}_\alpha),
\\
&&\Gamma(p\to K^+ \bar{\nu}) = \sum_{\alpha=e,\mu,\tau}\Gamma(p\to K^+ \bar{\nu}_\alpha),
\\
&&\Gamma(p\to \pi^0 \beta^+),
\\
&&\Gamma(p\to \eta \,\beta^+),
\\
&&\Gamma(p\to K^0 \beta^+), \ \ \ \ \  \ \ \ \ \  \ \ \ \ \  \ \ \ \  \ \ \ \ \ (\beta=e,\mu),
\eea
 which are suitable for the presentation because they are observable quantities.
 \\

The partial width of each mode is given by
\begin{align}
&\Gamma(p\to \pi^+\bar{\nu}_\beta)
\ = \ {\cal C}
\left\vert \beta_H(\mu_{\rm had})\frac{1}{f_\pi}
\left(1+D+F\right)C_{LL}^{d\beta \,ud}(\mu_{\rm had})\right\vert^2 \ \ \ \ \ (\beta=e,\mu),
\label{partial1}
\\
&\Gamma(p\to \pi^+\bar{\nu}_{\tau})
\ = \ {\cal C}
\left\vert \beta_H(\mu_{\rm had})\frac{1}{f_\pi}
\left(1+D+F\right)C_{LL}^{d\tau \,ud}(\mu_{\rm had})+\alpha_H(\mu_{\rm had})\frac{1}{f_\pi}C_{RL}^{ud \,\tau d}(\mu_{\rm had})
\right|^2,
\label{partial2}
\\
&\Gamma(p\to K^+\bar{\nu}_\beta)
\ = \ {\cal C}
\left\vert \beta_H(\mu_{\rm had})\frac{1}{f_\pi}
\left\{
\left(1+\frac{D}{3}+F\right)C_{LL}^{s\beta \,ud}(\mu_{\rm had})+
\frac{2D}{3}C_{LL}^{d\beta \,us}(\mu_{\rm had})\right\}\right\vert^2 \ \ \ \ \ (\beta=e,\mu),
\label{partial3}
\\
&\Gamma(p\to K^+\bar{\nu}_{\tau})
\ = \ {\cal C}
\left\vert \beta_H(\mu_{\rm had})\frac{1}{f_\pi}
\left\{
\left(1+\frac{D}{3}+F\right)C_{LL}^{s\tau \,ud}(\mu_{\rm had})+
\frac{2D}{3}C_{LL}^{d\tau \,us}(\mu_{\rm had})\right\}\right.
\nn\\
& \ \ \ \ \ \ \ \ \  \ \ \ \ \ \ \ \ \ \ \ \ \ \ \ \ \ \ \left.+\alpha_H(\mu_{\rm had})\frac{1}{f_\pi}\left\{
\left(1+\frac{D}{3}+F\right)C_{RL}^{ud \,\tau s}(\mu_{\rm had})+\frac{2D}{3}C_{RL}^{us \,\tau d}(\mu_{\rm had})
\right\}\right|^2,
\label{partial4}
\\
&\Gamma(p\to \pi^0\beta^+)
\ = \ {\cal C}
\left\vert \beta_H(\mu_{\rm had})\frac{1}{f_\pi}\frac{1}{\sqrt{2}}
\left(1+D+F\right)\ov{C}_{LL}^{u\beta \,ud}(\mu_{\rm had})
\right\vert^2
\label{partial5}
\\
&\Gamma(p\to \eta\,\beta^+)
\ = \ {\cal C}
\left\vert \beta_H(\mu_{\rm had})\frac{1}{f_\pi}\sqrt{\frac{3}{2}}
\left(1-\frac{D}{3}+F\right)\ov{C}_{LL}^{u\beta \,ud}(\mu_{\rm had})
\right\vert^2
\label{partial6}
\\
&\Gamma(p\to K^0\beta^+)
\ = \ {\cal C}
\left\vert \beta_H(\mu_{\rm had})\frac{1}{f_\pi}
\left(1-D+F\right)\ov{C}_{LL}^{u\beta \,us}(\mu_{\rm had})
\right\vert^2, \ \ \ \ \ \ \ \ \ \ \ \ \ (\beta=e,\mu),
\label{partial7}
\end{align}
  where ${\cal C} = \frac{m_N}{64\pi}\left(1-\frac{m_{K}^2}{m_N^2}\right)^2$, and
  $\alpha_H,\beta_H$ denote hadronic matrix elements, and $D,F$ are parameters of the baryon chiral Lagrangian.
The mass splittings among nucleons and hyperons are neglected.

In the evaluation of the proton decay partial widths,
 the baryon chiral Lagrangian parameters are given by $D=0.804$, $F=0.463$,
 and the hadronic matrix elements are taken from Ref.~\cite{Aoki:2017puj} as
 $\alpha_H(\mu_{\rm had})=-\beta_H(\mu_{\rm had})=-0.0144$~GeV$^3$ for $\mu_{\rm had}=2$~GeV.

In the calculation of $C_{LL\alpha}^0$, $C_{RL}^0$, defined in Eqs.~(\ref{cll0}),(\ref{crl0}),
 we assume two benchmark SUSY particle mass spectra.
In one spectrum, the pole masses and $\tan\beta$ satisfy
\bea
&&m_{\rm sfermion}=m_{H^0}=m_{H^\pm}=m_A=|M_{\widetilde{g}}|=|M_{\widetilde{W}}|=|\mu_h|=100~{\rm TeV},
\nn\\
&&\tan\beta = 5,
\label{massspectrum}
\eea
and in the other spectrum, they satisfy
\bea
&&m_{\rm sfermion}=m_{H^0}=m_{H^\pm}=m_A=|M_{\widetilde{g}}|=|M_{\widetilde{W}}|=|\mu_h|=30~{\rm TeV},
\nn\\
&&\tan\beta = 50,
\label{massspectrum2}
\eea
 where all the sfermions are mass-degenerate.
The relative phase between $M_{\widetilde{W}}$ and $\mu_h$, which determines the relative phase between $C_{LL\alpha}^0$ and $C_{RL}^0$, is varied randomly.
The first spectrum represents the case with low $\tan\beta$ and the second spectrum the case with high $\tan\beta$.
In both spectra, 
 the correct electroweak symmetry breaking and the values of $m_{H^0}$, $m_{H^\pm}$, $m_A$
 can be achieved by a fine-tuning of soft SUSY breaking parameters $m_{H_u}^2$, $m_{H_d}^2$, $B\mu$
 (since $m_{H^0}$, $m_{H^\pm}$, $m_A$ are much above the electroweak scale, 
  the equality $m_{H^0}=m_{H^\pm}=m_A$ holds to a good precision).
Both spectra satisfy the 1-loop matching condition~\cite{Ellis:2017erg} for the SM Higgs quartic coupling
 around scale $\mu=m_{\rm sfermion}$ (with vanishing $A$-terms) and thus can realize the correct Higgs particle mass 125~GeV.
For the first spectrum, the model evades the current experimental bounds on proton decay, including the 90\% CL bound on the $p\to K^+\bar{\nu}$ mode,
 $1/\Gamma(p\to K^+\bar{\nu})>5.9\times10^{33}$~yrs~\cite{Abe:2014mwa}, when $({\cal M}_{H_C}^{-1})_{11} \lesssim 0.02/M_P$
  (slightly smaller than the estimate after Eq.~(\ref{mhcinverse})).
For the second spectrum, the model evades the current experimental bounds when $({\cal M}_{H_C}^{-1})_{11} \lesssim 2/M_P$.
The SUSY particle masses can be made smaller and be at TeV scale 
 if we set $({\cal M}_{H_C}^{-1})_{11} \ll 1/M_P$ by taking $|\kappa|\ll1$.
However, since such small $\kappa$ is not natural, we adhere to the relation $({\cal M}_{H_C}^{-1})_{11} \sim 1/M_P$
 and assume $O(10)$ to 100~TeV SUSY particle mass spectra.
We comment on the unification of the $SU(3)_C$ and $SU(2)_L$ gauge couplings $g_3,~g_2$
 under the above SUSY particle mass spectrum.
Let us focus on the case with $|\lambda\lambda'|\sim 1$ and where the GUT-breaking VEVs are smaller than the Planck scale.
We note that in this case the colored Higgs masses $M_{H_{C1}},\,M_{H_{C2}}$ originate mostly from the GUT-breaking VEVs and satisfy 
 $M_{H_{C1}}M_{H_{C2}}=|\lambda\lambda' \langle N_H^c\rangle\langle \ov{N}_H^c\rangle|$.
\footnote{
To see this, note that the vacuum condition Eq.~(\ref{vev}) gives $|\langle N_H^c\rangle|=|\langle \ov{N}_H^c\rangle|\gg M$.
Then the mass matrix Eq.~(\ref{coloredhiggsmass}) leads to 
 $M_{H_{C1}}M_{H_{C2}}=|\lambda\lambda' \langle N_H^c\rangle\langle \ov{N}_H^c\rangle|$.
 }
On the other hand, the GUT gauge boson mass $M_G$ satisfies 
 $M_G^2=g_{23}^2|\langle N_H^c\rangle\langle \ov{N}_H^c\rangle|$ ($g_{23}$ denotes the unified gauge coupling).
Then we get $M_{H_{C1}}M_{H_{C2}}\sim M_G^2$, and we can use the 1-loop unification condition,
\bea
\frac{1}{g_3^2(\mu)} - \frac{1}{g_2^2(\mu)} \ = \ 2\log\frac{\mu^2}{M_{H_{C1}}M_{H_{C2}}} + 4\log\frac{\mu}{M_G}.
\eea
Numerically evaluating the condition, we obtain
\bea
M_G^2M_{H_{C1}}M_{H_{C2}} \ = \ (1\cdot10^{16}~{\rm GeV})^4
\eea
 with negligible dependence on $\tan\beta$. The unification of $g_3,~g_2$ constrains the colored Higgs masses and GUT gauge boson mass as above.

The RG corrections for the Wilson coefficients in MSSM and SM are calculated 
 by using 1-loop RG equations in Ref.~\cite{Hisano:1992jj,Goto:1998qg}.
We fix $\mu_{H_C}=2\cdot10^{16}$~GeV and $\mu_{\rm SUSY}=m_{\rm sfermion}$.

The right-hand sides of Eqs.~(\ref{cll1-2})-(\ref{crl2-2}) are evaluated by solving the 2-loop RG equations of the Yukawa couplings
 in SM and MSSM with the above SUSY particle mass spectrum.
The input values of the RG equations are given in terms of quark and lepton masses and Wolfenstein parameters,
 and are taken from the central values of the following experimental data:
The isospin-averaged quark mass and strange quark mass in $\ov{{\rm MS}}$ scheme are obtained from lattice calculations in
 Refs.~\cite{lattice1,lattice2,lattice3,lattice4,lattice5,lattice6} as
 $\frac{1}{2}(m_u+m_d)(2~{\rm GeV})=3.373(80)~{\rm MeV}$ and $m_s(2~{\rm GeV})=92.0(2.1)~{\rm MeV}$.
The up and down quark mass ratio is obtained from an estimate in Ref.~\cite{latticereview} as $m_u/m_d=0.46(3)$.
The $\ov{{\rm MS}}$ charm and bottom quark masses are obtained from QCD sum rule calculations in Ref.~\cite{cb} as
  $m_c(3~{\rm GeV})=0.986-9(\alpha_s^{(5)}(M_Z)-0.1189)/0.002\pm0.010~{\rm GeV}$
  and $m_b(m_b)=4.163+7(\alpha_s^{(5)}(M_Z)-0.1189)/0.002\pm0.014~{\rm GeV}$.
The top quark pole mass is obtained from the latest measurement of the CMS Collaboration~\cite{Sirunyan:2019zvx}
 as $170.5\pm0.8$~GeV.
The values of the Wolfenstein parameters are taken from the CKM fitter result~\cite{ckmfitter}.
For the QCD and QED gauge couplings, we use $\alpha_s^{(5)}(M_Z)=0.1181$ and $\alpha^{(5)}(M_Z)=1/127.95$.
For the lepton and W, Z, Higgs pole masses, we use the values in Particle Data Group~\cite{ParticleDataGroup:2020ssz}.

We comment on the impact of the choice of the benchmark spectrum in Eq.~(\ref{massspectrum}).
If the spectrum deviates from Eq.~(\ref{massspectrum}) and 
 the masses of isospin-doublet sleptons $\tilde{\ell}^\alpha$ are split, this gives rise to a splitting in $C_{LL\alpha}^0$'s,
 which affects proton decay partial width ratios.
However, since the right-hand sides of Eqs.~(\ref{cll1-2})-(\ref{crl2-2}) have a large hierarchy,
 the possible splitting in $C_{LL\alpha}^0$'s has only a minor impact on proton decay partial width ratios.
Likewise, a splitting in the 1st and 2nd generation isospin-doublet squark masses does not change the result significantly.
\\

The unknown unitary matrix $U$ is varied with the Haar measure given by~\cite{Haba:2000be,Lu:2014cla}
\bea
&&U=e^{i\,\eta}e^{i\,\omega_1\lambda_3+i\,\omega_2\lambda_8} \begin{pmatrix} % or pmatrix or bmatrix or Bmatrix or ...
      1 & 0 & 0 \\
      0 & \cos\theta_{23} & \sin\theta_{23} \\
      0 & -\sin\theta_{23} &  \cos\theta_{23} \\
   \end{pmatrix}
   \begin{pmatrix} % or pmatrix or bmatrix or Bmatrix or ...
      \cos\theta_{13} & 0 & \sin\theta_{23}e^{-i\,\delta} \\
      0 & 1 & 0 \\
      -\sin\theta_{23}e^{i\,\delta} & 0 &  \cos\theta_{13} \\
   \end{pmatrix}
  \nn\\
  &&\times
      \begin{pmatrix} % or pmatrix or bmatrix or Bmatrix or ...
      \cos\theta_{12} & \sin\theta_{12} & 0 \\
      -\sin\theta_{12} & \cos\theta_{12} & 0 \\
      0 & 0 & 1\\
   \end{pmatrix}e^{i\,\chi_1\lambda_3+i\,\chi_2\lambda_8},
\\
&&{\rm d}U={\rm d}\sin^2\theta_{23}\ {\rm d}\cos^4\theta_{13}\ {\rm d}\sin^2\theta_{12}\ 
{\rm d}\eta\ {\rm d}\omega_1\ {\rm d}\omega_2\ {\rm d}\delta\,{\rm d}\chi_1\ {\rm d}\chi_2,
\eea
 where $\lambda_3={\rm diag}(1,-1,0)$, $\lambda_8={\rm diag}(1,1,-2)/\sqrt{3}$.
The use of the Haar measure is justifiable because we have no information on the flavor basis of ${\bf \bar{5}}_{-3}^i$'s
 and the Haar measure is invariant under an arbitrary unitary transformation on the basis.
The unknown phases $\phi_1,\phi_2,\phi_3$ and the relative phase between $M_{\widetilde{W}}$ and $\mu_h$
 are varied with the flat distribution.
\\

\begin{figure}[H]
\begin{center}
\includegraphics[width=80mm]{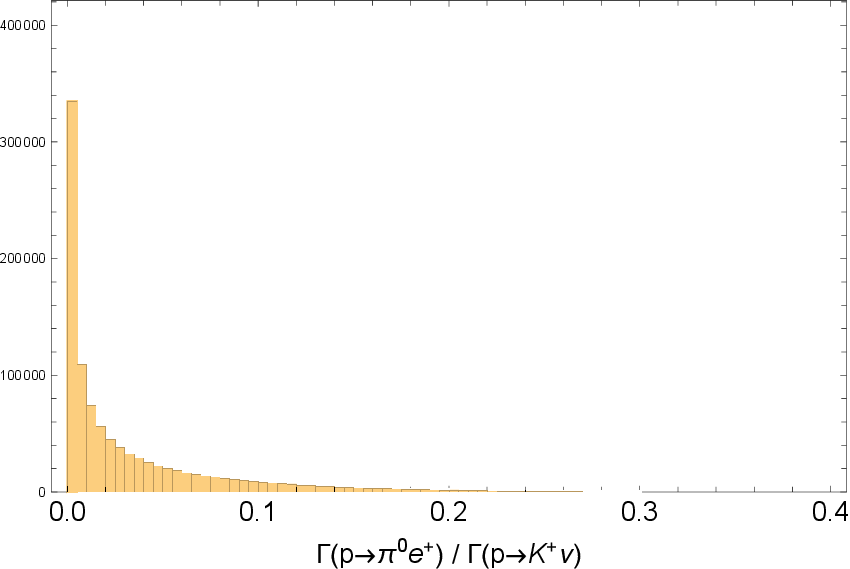}
\\
\includegraphics[width=80mm]{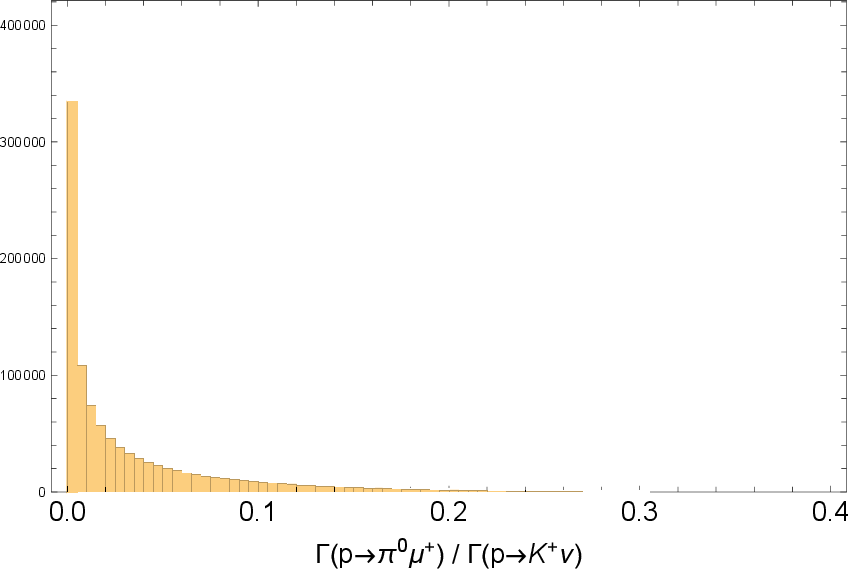}    \ \ \ \ \ \                          \includegraphics[width=80mm]{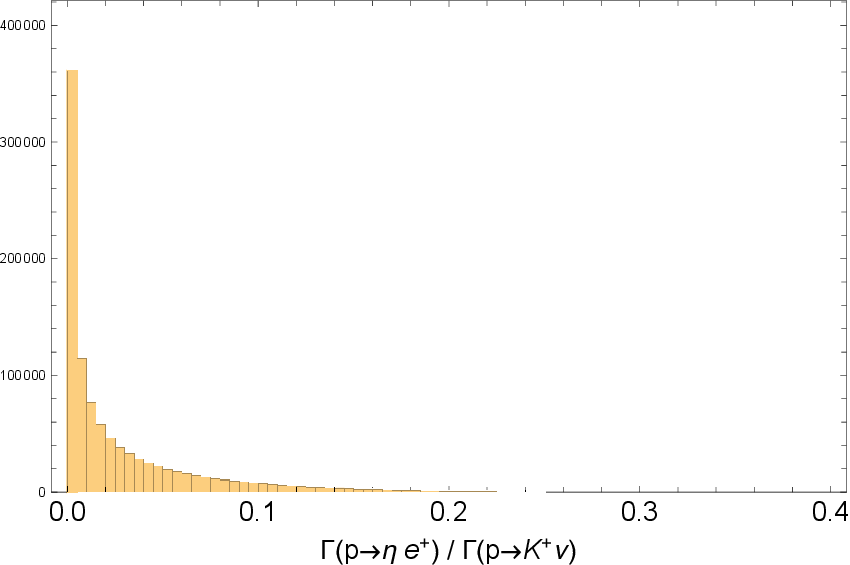}
\\
\includegraphics[width=80mm]{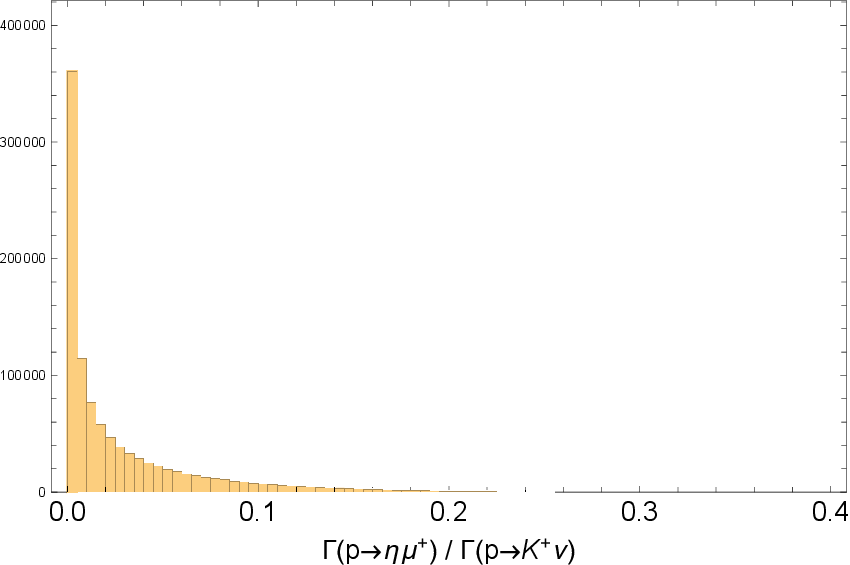}    \ \ \ \ \ \                          \includegraphics[width=80mm]{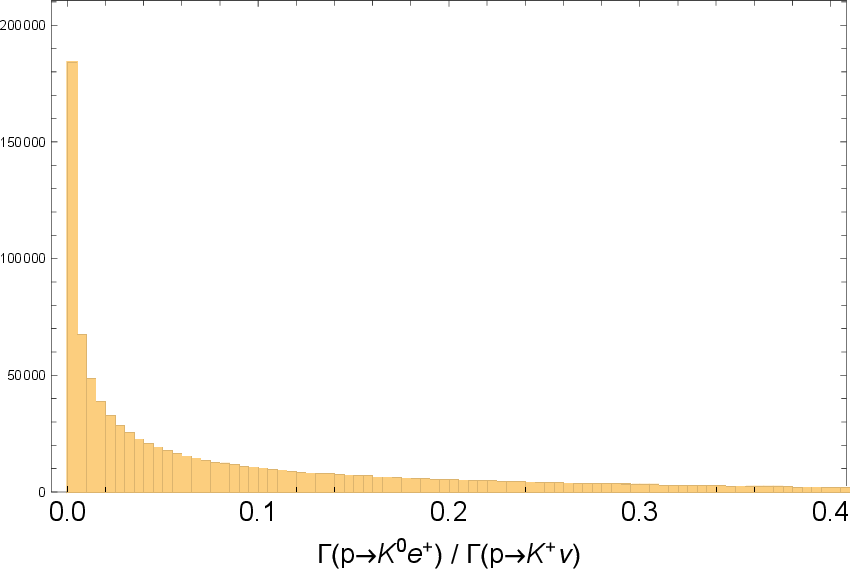}
\\
\includegraphics[width=80mm]{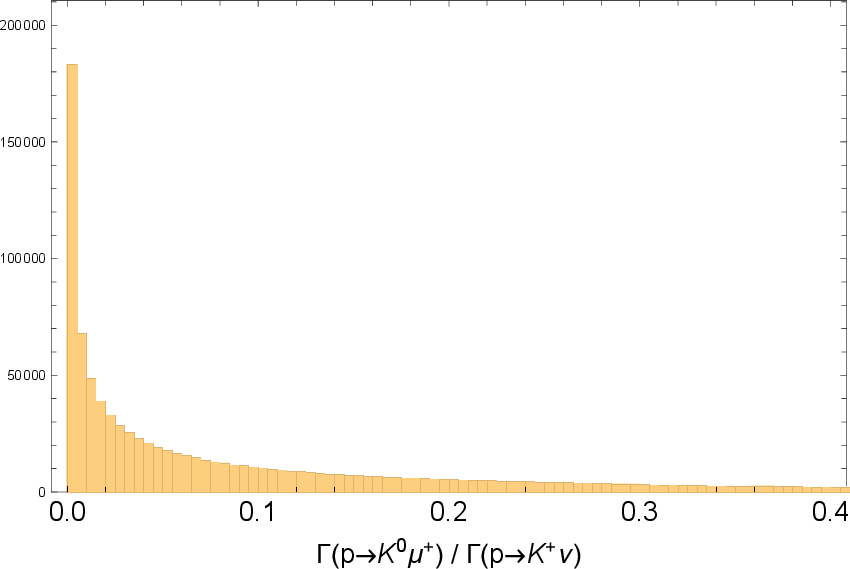}    \ \ \ \ \ \                          \includegraphics[width=80mm]{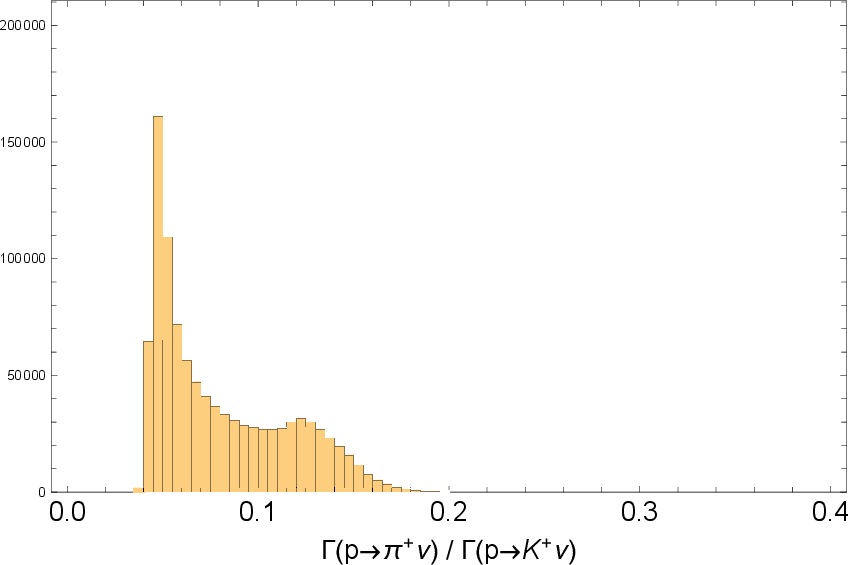}
\caption{
Distributions of proton decay partial width ratios corresponding to randomly varied values of the unknown unitary matrix $U$, unknown phases $\phi_1,\phi_2,\phi_3$,
 and relative phase between $M_{\widetilde{W}}$ and $\mu_h$.
The vertical axis is linear and in arbitrary units. Here we assume $\tan\beta=5$.
}
\label{tanbeta5}
\end{center}
\end{figure}

\begin{figure}[H]
\begin{center}
\includegraphics[width=80mm]{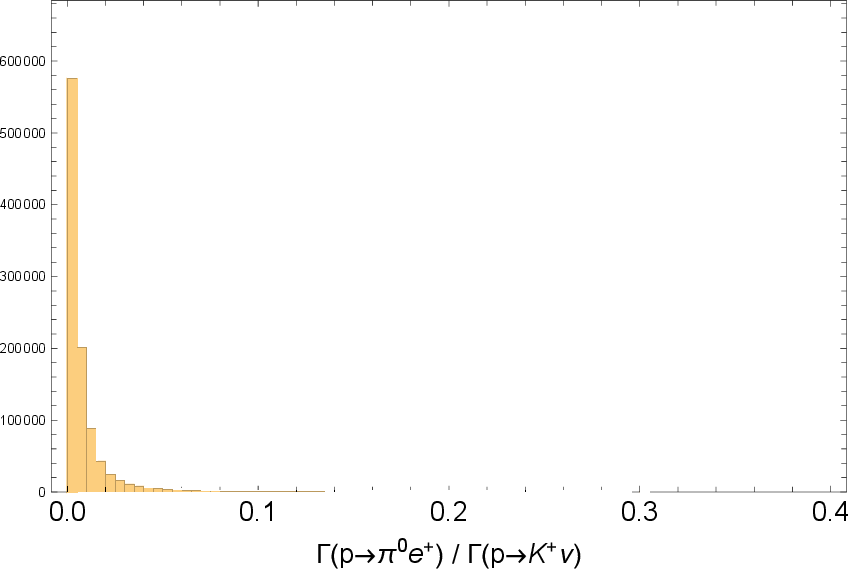}
\\
\includegraphics[width=80mm]{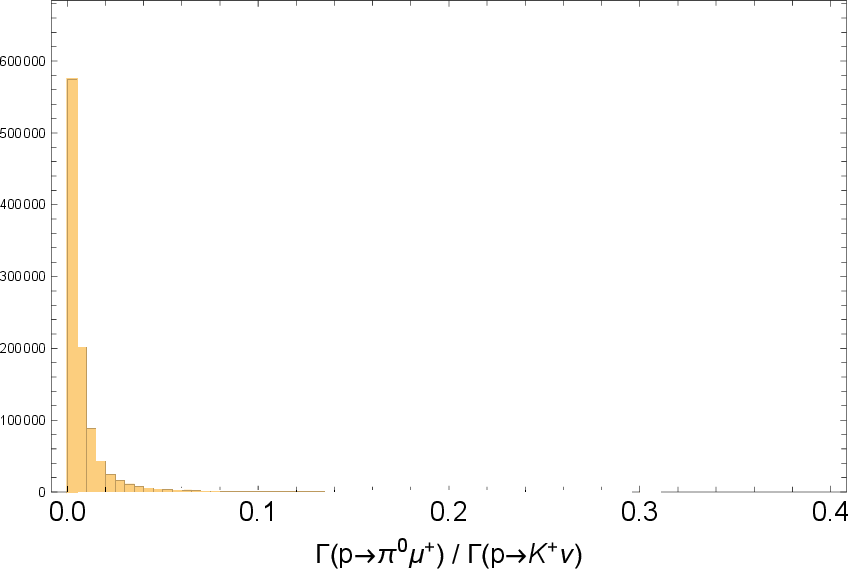}    \ \ \ \ \ \                          \includegraphics[width=80mm]{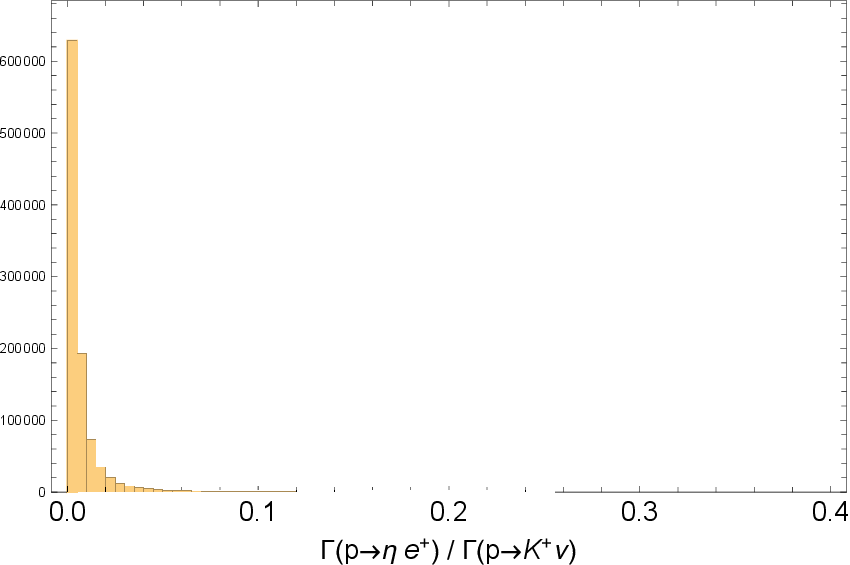}
\\
\includegraphics[width=80mm]{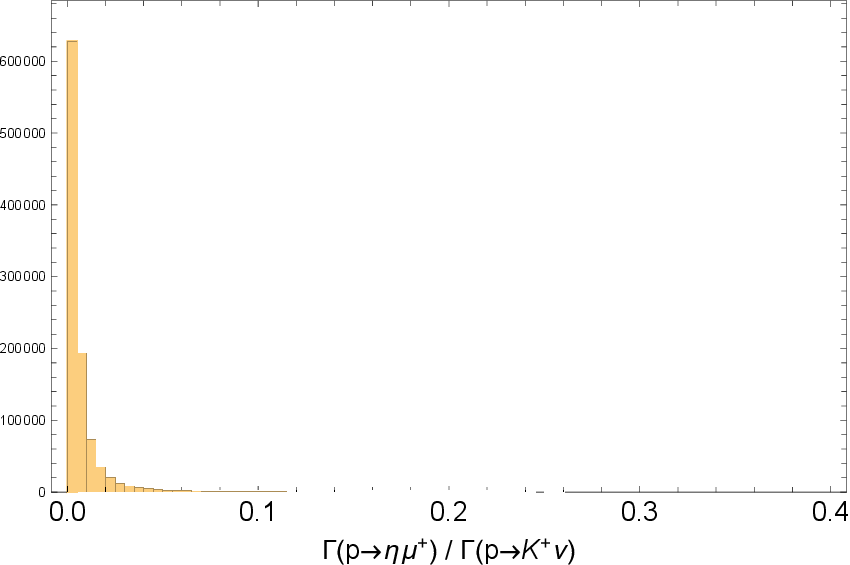}    \ \ \ \ \ \                          \includegraphics[width=80mm]{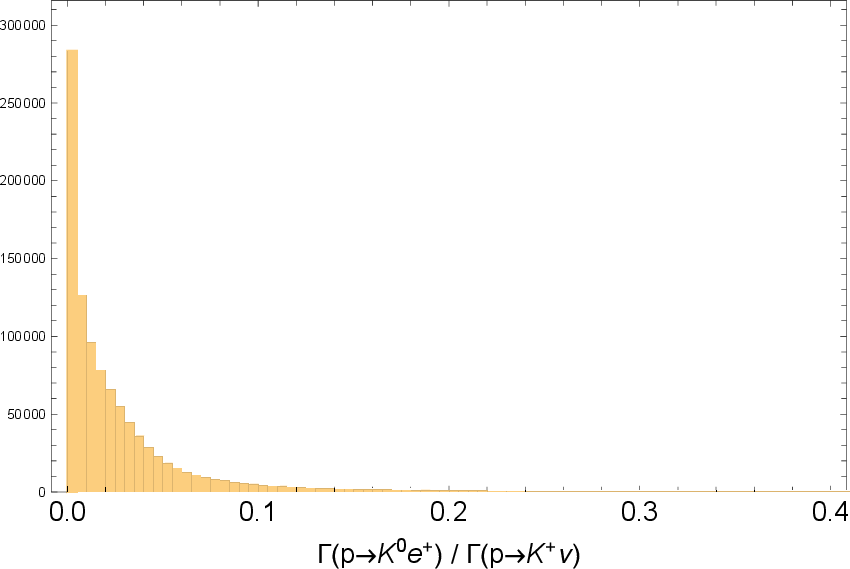}
\\
\includegraphics[width=80mm]{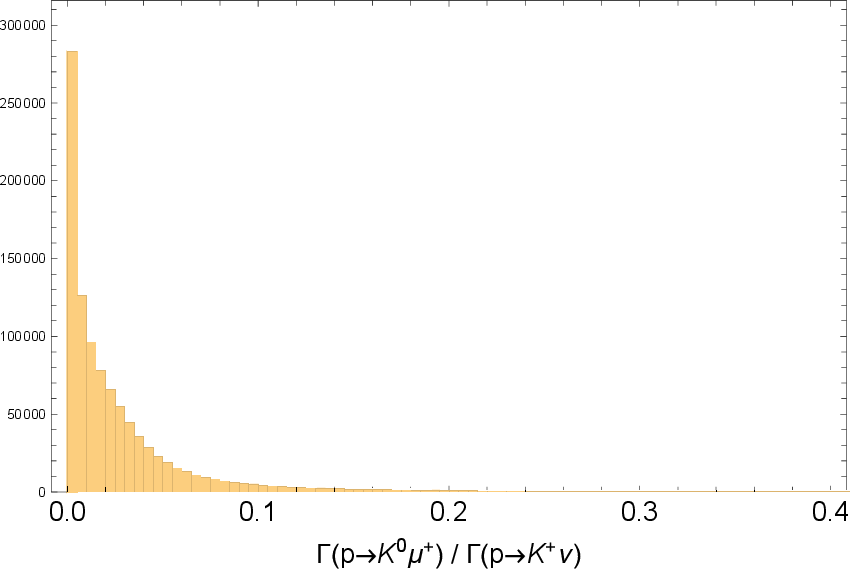}    \ \ \ \ \ \                          \includegraphics[width=80mm]{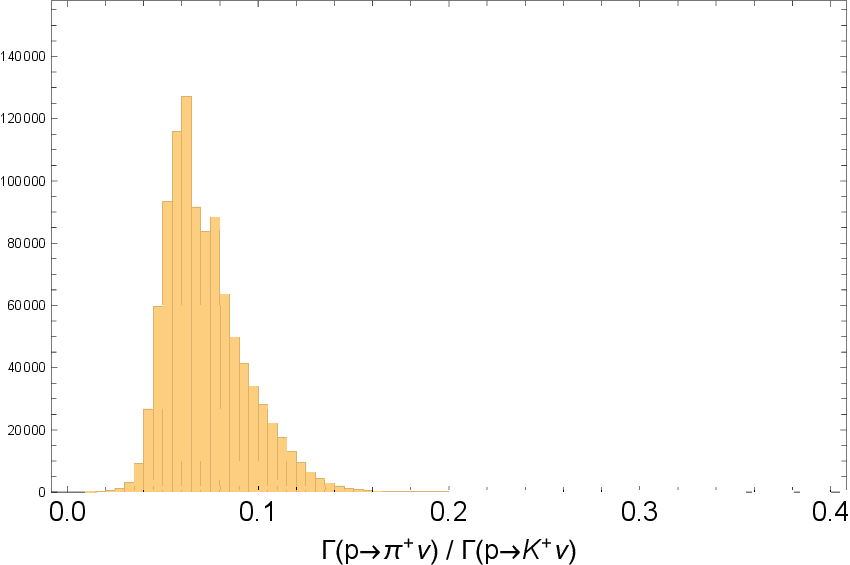}
\caption{
Same as Fig.~\ref{tanbeta5} except that we assume $\tan\beta=50$.
}
\label{tanbeta50}
\end{center}
\end{figure}

We present the result of the numerical analysis.
Since $\Gamma(p\to K^+ \bar{\nu})$ is the largest partial width in the entire parameter space,
 the phenomenologically most meaningful quantities are the ratios of $\Gamma(p\to K^+ \bar{\nu})$ and the other partial widths.
Therefore, we show the distributions of
\bea
\frac{\Gamma(p\to \pi^+ \bar{\nu})}{\Gamma(p\to K^+ \bar{\nu})}, \ \ \
\frac{\Gamma(p\to \pi^0 \beta^+)}{\Gamma(p\to K^+ \bar{\nu})}, \ \ \ 
\frac{\Gamma(p\to \eta \beta^+)}{\Gamma(p\to K^+ \bar{\nu})}, \ \ \
\frac{\Gamma(p\to K^0 \beta^+)}{\Gamma(p\to K^+ \bar{\nu})} \ \ \ (\beta=e,\mu)
\eea
 corresponding to randomly varied values of the unknown unitary matrix $U$, unknown phases $\phi_1,\phi_2,\phi_3$,
 and relative phase between $M_{\widetilde{W}}$ and $\mu_h$.
Figs.~\ref{tanbeta5},~\ref{tanbeta50} are the distributions for $\tan\beta=5$, 50, respectively.

\newpage
 \thispagestyle{empty} 

From Figs.~\ref{tanbeta5},\ref{tanbeta50}, we find that for low $\tan\beta$ such as $\tan\beta=5$,
 the ratio $\frac{\Gamma(p\to\pi^+\bar{\nu})}{\Gamma(p\to K^+\bar{\nu})}$ can be in the range 0.1-0.2 with $O(0.1)$ probability.
Therefore, future sensitivity study on the $p\to\pi^+\bar{\nu}$ mode should be performed seriously,
 to examine the possibility of observing both decay modes.
The other partial width ratios are mostly below 0.05 for both low and high $\tan\beta$.
However, for low $\tan\beta$ such as $\tan\beta=5$, there is a non-negligible probability that 
 $\frac{\Gamma(p\to \pi^0 \beta^+)}{\Gamma(p\to K^+ \bar{\nu})}$ and
 $\frac{\Gamma(p\to \eta \beta^+)}{\Gamma(p\to K^+ \bar{\nu})}$ are in the range 0.05-0.1.
In this case, the current bound on $\Gamma(p\to K^+\bar{\nu})$~\cite{Abe:2014mwa}
 and the future sensitivity reach for $\Gamma(p\to \pi^0 \beta^+)$~\cite{Hyper-Kamiokande:2018ofw}
 imply that $p\to \pi^0 \beta^+$, along with $p\to K^+ \bar{\nu}$, can be discovered.
Also, the distributions of $\frac{\Gamma(p\to K^0 \beta^+)}{\Gamma(p\to K^+ \bar{\nu})}$ show a long tail above 0.05,
 which indicates that there is a non-zero probability that $p\to K^0 \beta^+$ can be discovered with a large rate
 with $\frac{\Gamma(p\to K^0 \beta^+)}{\Gamma(p\to K^+ \bar{\nu})}=O(0.1)$.
\\

\section{Summary}
\label{summary}

We have studied dimension-five proton decay in a model based on the flipped $SU(5)$ GUT.
In the model, 
 the GUT-breaking ${\bf10},\,{\bf\ov{10}}$ fields have a GUT-scale mass term and 
 gain VEVs through operators suppressed by the Planck scale.
This structure induces an effective mass term not much smaller than the GUT scale between the color triplets in the ${\bf 5}$,~${\bf \bar{5}}$ Higgs fields.
This mass term gives rise to observable dimension-five proton decay, 
 and at the same time achieves moderate suppression on dimension-five proton decay amplitudes,
 which is estimated to be 0.01 if the coefficients in the superpotential Eq.~(\ref{superpotential}) satisfies 
 $|\gamma_1/(\lambda\lambda')|=1$.

We have investigated the flavor structure of the Wilson coefficients of the operators contributing to dimension-five proton decay,
 and expressed them in terms of diagonalized Yukawa couplings and CKM matrix components in MSSM
 plus an unknown unitary matrix $U$ and unknown phases.
We have numerically evaluated the Wilson coefficients by randomly varying $U$ and the unknown phases,
 and presented the result in the form of the distributions of the partial width ratios of various proton decay modes for a benchmark SUSY particle spectrum.
We have found that the ratio
 $\frac{\Gamma(p\to\pi^+\bar{\nu})}{\Gamma(p\to K^+\bar{\nu})}$ can be in the range 0.1-0.2 with $O(0.1)$ probability for low $\tan\beta$ such as $\tan\beta=5$.
Also, for such low $\tan\beta$, it is possible that $\frac{\Gamma(p\to \pi^0 \beta^+)}{\Gamma(p\to K^+ \bar{\nu})}$ and
 $\frac{\Gamma(p\to \eta \beta^+)}{\Gamma(p\to K^+ \bar{\nu})}$ are in the range 0.05-0.1,
 and there is a non-zero probability that $\frac{\Gamma(p\to K^0 \beta^+)}{\Gamma(p\to K^+ \bar{\nu})}=O(0.1)$.
\\

%%%%%%%%%%%%%%%
\section*{Acknowledgement}
This work is partially supported by Scientific Grants by the Ministry of Education, Culture, Sports, Science and Technology of Japan
 Nos.~17K05415 and 21H000761 (N.H.) and No.~19K147101 (T.Y.).
\\

%%%%%%%%%%%%%%%
%%% References %%%
%%%%%%%%%%%%%%%

\end{document}